\documentclass[12pt,a4paper]{article}

\usepackage[T1]{fontenc}
\usepackage[hmarginratio=1:1,top=32mm,columnsep=20pt]{geometry}
\usepackage{mathdots}

\usepackage{color}
\definecolor{dark-gray}{gray}{0.20}
\definecolor{gray}{gray}{0.30}
\definecolor{light-gray}{gray}{0.80}
\definecolor{dark-red}{rgb}{0.7,0,0}
\definecolor{dark-green}{rgb}{0.1,0.4,0}
\definecolor{dark-blue}{rgb}{0.3,0.3,0.7}
\definecolor{light-blue}{rgb}{0.8,0.8,1}

\usepackage{cite}
\usepackage{hyperref}
\hypersetup{
	colorlinks=true,
	linkcolor=dark-blue,
	citecolor=dark-red,
	urlcolor=dark-green,
	linktoc=page
}

  \usepackage{a4wide}
  \usepackage{latexsym}
  \usepackage{epsf}
  \usepackage{bbm}
  \usepackage{graphicx}
  \usepackage{amsmath,amssymb,amsthm}
  \usepackage{verbatim}

\newcommand{\e}{\mathrm{e}}

\newcommand{\bbm}{\left(\begin{matrix}}
\newcommand{\ebm}{\end{matrix}\right)}
\newcommand{\bea}{\begin{eqnarray}}
\newcommand{\eea}{\end{eqnarray}}
\newcommand{\be}{\begin{equation}}
\newcommand{\ee}{\end{equation}}

\renewcommand{\d}{\textrm{d}}

\newcommand{\SL}{\mathop{\rm SL}}
\newcommand{\GL}{\mathop{\rm GL}}
\newcommand{\SO}{\mathop{\rm SO}}

\newcommand{\SU}{\mathop{\rm SU}}
\newcommand{\U}{\mathop{\rm U}}

\begin{document}

\numberwithin{equation}{section}

\begin{center}

{\LARGE {\bf Instantons from geodesics in AdS moduli spaces}}  \\

\vspace{1.5 cm} {\large  Daniele Ruggeri $^{a,b}$, Mario Trigiante$^{a,b}$ and  Thomas Van Riet$^c$ }\footnote{{ \upshape\ttfamily daniele.rug@gmail.com, mario.trigiante@polito.it,  thomas.vanriet @fys.kuleuven.be
 } }\\
\vspace{0.5 cm}  \vspace{.15 cm} { ${}^a$Department of Applied Science and Technology, Politecnico di Torino, \\ C.so Duca degli Abruzzi, 24, I-10129 Torino, Italy\\ \vspace{.15 cm}
${}^b $ Istituto Nazionale di Fisica Nucleare (INFN)\\  Sezione di Torino, via P.
Giuria 1, Turin, Italy \\ \vspace{.15 cm}
${}^c$Instituut voor Theoretische Fysica, K.U. Leuven,\\
Celestijnenlaan 200D B-3001 Leuven, Belgium
}

\vspace{2cm}

{\bf Abstract}
\end{center}

{\small We investigate supergravity instantons in Euclidean $\rm AdS_5\times S^5/\mathbb{Z}_k$. These solutions are expected to be dual to instantons of $\mathcal{N}=2$ quiver gauge theories. On the supergravity side the (extremal) instanton solutions are neatly described by the (lightlike) geodesics on the AdS moduli space for which we find the explicit expression and compute the on-shell actions in terms of the quantised charges. The lightlike geodesics fall into two categories depending on the degree of nilpotency of the Noether charge matrix carried by the geodesic: For degree 2 the instantons preserve 8 supercharges and for degree 3 they are non-SUSY.   We expect that these findings should apply to more general situations in the sense that there is a map between geodesics on moduli-spaces of Euclidean AdS vacua and instantons with holographic counterparts. 

\setcounter{tocdepth}{2}
\newpage
\tableofcontents
\newpage

\section{Introduction}
Since the birth of the AdS/CFT correspondence it has been natural to identify the gauge theory interpretation, if any, of supergravity solutions that are asymptotically AdS. For the $\rm AdS_5 \times S^5$ vacuum of IIB supergravity, a particularly simple class of deformations can be found by switching on the dilaton $\phi$ and the RR axion $\chi$. For instance, the D-instanton in Euclidean $\rm AdS_5 \times S^5$ is of this type and can be regarded as the near horizon of a $D3/D(-1)$-intersection \cite{Chu:1998in}. This solution is suggested to be the dual to specific (supersymmetric) instantons in $\mathcal{N}=4$ SYM theory \cite{Banks:1998nr}. This conjecture has survived many non-trivial checks \cite{Dorey:1998xe, Dorey:1998qh, Dorey:1999pd, Green:2002vf}, which were reviewed in \cite{Belitsky:2000ws}. The simplest checks involved the matching of the on-shell actions, the charges, the supersymmetries, the moduli-spaces as well as the computation of the holographic one-point functions $\langle \text{Tr}F^2\rangle$ and $\langle \text{Tr}F\tilde{F}\rangle$, see \cite{Balasubramanian:1998de, Chu:1998in, Bianchi:1998nk, Kogan:1998re}.

One can wonder how much of this matching is fixed by the conformal supersymmetry. Hence it is interesting to break supersymmetry and study non-SUSY instantons. This is much easier from the gravity side than from the field theory side, and simple non-extremal instanton solutions in Euclidean $\rm AdS_5 \times S^5$ are indeed easily constructed \cite{Bergshoeff:2005zf} (see also \cite{Gutperle:2002km}). They can be organised according to the ratio of their charge $Q$ to the on-shell action $S$: the solution is  named `over-extremal' when $S<Q$, `under-extremal' when $S>Q$. The geometry of the over-extremal solution is a two-sided AdS wormhole.  These wormholes are examples of the well-know  Euclidean axionic wormholes of \cite{Giddings:1987cg, Coleman:1988cy, Lavrelashvili:1987jg} whose understanding is still unclear. A problematic feature of the two-sided axion wormhole in $\rm AdS_5\times S^5$ is an unphysical singularity in the axion-dilaton scalar profiles (that cancel out in the energy-momentum tensor). Indeed, a naive computation \cite{Bergshoeff:2005zf} that ignores the subtleties of having two boundaries, indicates that the would-be solution violates the BPS bound in the dual field theory since:
\be\label{inconsistent}
\langle \text{Tr}F^2\rangle <\langle \text{Tr}F\tilde{F}\rangle\,,
\ee
which is inconsistent. It is natural to conjecture that the inconsistent operator values (\ref{inconsistent}) are a consequence of the singularity in the scalar fields and hence the solution is discarded.

The under-extremal solutions have a ``spike-like'' singularity in the bulk, which is identified with the position of the D-instanton. Such solutions, when embedded in flat space, can be seen as non-extremal black holes reduced over time, and hence might be physical. The holographic one-point functions now imply that these could be dual to non-self dual YM instantons satisfying:
\be
\langle \text{Tr}F^2\rangle >\langle \text{Tr}F\tilde{F}\rangle\,.
\ee
A concrete suggestion for the holographic dual description was made in \cite{Bergshoeff:2004pg}: The instantons correspond to the addition of an anti-self-dual $\rm SU(2)$ instanton $\overline{A}_{\mu}$  in the colormatrix which has for the rest order $N$ self-dual $\rm SU(2)$ instantons ${A}_{\mu}$ on mutually commuting blocks as follows:
\be
A_{\mu}^{\rm SU(N)}=\begin{pmatrix}
A_{\mu}^{\rm SU(2)} & 0 & \ldots & 0\\
 0 & A_{\mu}^{\rm SU(2)} &  & 0\\
 \vdots  &  & \ddots  & \\
0   & &  & \overline{A}_{\mu}^{\rm SU(2)}
\end{pmatrix}\,.
\ee

For all solutions described above the axion and dilaton only depend on the $\rm AdS_5$ coordinates such that they can be derived from the following simple truncation of 5-dimensional maximal $\SO(6)$-gauged supergravity:
\be\label{axion-dilaton}
S = -\frac{1}{2\kappa_5^2} \int \sqrt{|g_5|}\Bigl(R_5 - \tfrac{1}{2}(\partial\phi)^2 - \tfrac{1}{2}\epsilon\e^{b\phi}(\partial\chi)^2 - \Lambda \Bigr)\,,
\ee
where $\Lambda<0$ is the cosmological constant and $\epsilon = +1$ in Lorentzian signature but $\epsilon =-1$ in Euclidean signature. The dilaton coupling $b$ is furthermore fixed to be $b=2$. It was pointed out in \cite{ArkaniHamed:2007js} that when $b\leq \sqrt{3/2}$ the Euclidean wormholes would have a regular axion-dilaton profile. Only if such a dilaton coupling could be found in string theory (or any other UV complete gravity theory) should one try to understand the meaning of these wormholes. It was furthermore suggested in \cite{ArkaniHamed:2007js} that such small dilaton couplings $b$ could be obtained in theories with more involved sigma models that contain multiple axion-dilaton pairs. Indeed a concrete example was recently found in \cite{Hertog:2017owm}. It turns out sufficient to consider $\rm AdS_5\times S^5/\mathbb{Z}_k$ with $k>1$. Then the AdS moduli space $\mathcal{M}_{\text{moduli}}$ has been computed to be $\mathcal{M}_{\text{moduli}}=\SU(1,k)/{\rm S[U(1)\times U(k)]}$ \cite{Corrado:2002wx, Louis:2015dca}. This means that there should be a consistent truncation down to the following action
\be\label{5Daction}
S = -\frac{1}{2\kappa_5^2} \int \sqrt{|g_5|}\Bigl(R_5 - \tfrac{1}{2}G_{ij}\partial\Phi^I\partial\Phi^J  - \Lambda \Bigr)\,,
\ee
with the $\Phi^I$ coordinates on $\mathcal{M}_{\text{moduli}}$ and $G_{IJ}$ its canonical metric. In Euclidean signature one instead finds a Wick-rotated version of the sigma model. The Wick-rotation of the sigma model is neither unique nor fixed by Euclidean supersymmetry, but given the higher-dimensional and holographic interpretation of the scalars it was shown to be \cite{Hertog:2017owm}:
\be
\text{Wick-rotation:}\quad\frac{\SU(1,k)}{{\rm S[U(1)\times U(k)]}} \Longrightarrow \frac{\SL(k+1,\mathbb{R})}{\GL(k,\mathbb{R})}\,.
\ee
This coset allows then consistent truncations to axion-dilaton Lagrangians of the form (\ref{axion-dilaton}) with $b=1$ giving regular axion wormholes \cite{Hertog:2017owm}. This prompts the question of their holographic meaning, which is still unclear, see for instance \cite{ArkaniHamed:2007js, Maldacena:2004rf}. Given this explicit embedding in string theory it should be possible to settle the question whether these wormholes contribute as saddle points in the path integral \cite{inprogress}.

We consider it natural to first understand the extremal instantons in $\rm AdS_5\times S^5/\mathbb{Z}_k$ before the more intricate cases of under- and over-extremal solutions. The goal of this paper is therefore:
\begin{enumerate} \item To construct all solutions explicitly, by solving the geodesic equations on $\frac{\SL(k+1,\mathbb{R})}{\GL(k,\mathbb{R})}$; \item To study the supersymmetry properties of the extremal solutions; \item To compute the on-shell action in terms of quantised charges.\end{enumerate}

Remarkably, it is possible to find the closed expression for all geodesics for arbitrary $k$, in contrast with earlier studies, in the context of black holes, where the expressions for the scalars become highly complicated and are easily filling several pages for a single scalar field.

The rest of this paper is organised as follows. In the next section \ref{sec:general} we explain the exact relation between geodesics on the moduli-space and instanton solutions and we provide the explicit solutions in the case of $\rm AdS_5\times S^5/\mathbb{Z}_k$. In section \ref{sec:on-shell} we compute the on-shell action of the instantons by first Hodge dualising all axionic scalars. The supersymmetry analysis of the extremal solutions is discussed in section \ref{sec:susy}. A brief discussion of the non-extremal solutions is given in section \ref{sec:non-extremal} and a summary of our results with a discussion on future applications can be found in section \ref{sec:discussion}. We have also added various appendices with technicalities required to carry out the computations.

\section{Instantons on $\rm AdS_5\times S^5/\mathbb{Z}_k$} \label{sec:general}
\subsection{General framework}
As explained in the introduction, the instanton solutions in Euclidean $\rm AdS_5\times S^5/\mathbb{Z}_k$ are expected to be solutions of 5D gauged supergravity obtained from compactifying Euclidean IIB supergravity on $\rm S^5/\mathbb{Z}_k$. The description of that gauged supergravity can be found in \cite{Corrado:2002wx}. The proof that maximal $D=5$ supergravity with gauge group ${\rm SO}(6)$ is a consistent truncation of Type IIB theory on $\rm AdS_5\times S^5$ was recently given in \cite{Ciceri:2014wya,Baguet:2015sma}. As far as the orbifolded case is concerned, the corresponding five-dimensional description in terms of a suitable gauged half-maximal supergravity is only conjectured.  Here we only need the consistency of the truncation down to the moduli space, i.e. the truncation to the exactly massless sector of the gauged supergravity. Within that truncation, the bosonic 5d action is given by (\ref{5Daction})
\be
S = -\frac{1}{2\kappa^2_5}\int \sqrt{|g_5|}\Bigl(R_5 - \tfrac{1}{2}G_{IJ}\partial\Phi^I\partial\Phi^J  - \Lambda \Bigr)\,,
\ee
where $G_{IJ}$ is the metric on the moduli space $\mathcal{M}_{\text{moduli}}$ and $\Lambda = -12/\ell^2$. In Lorentzian signature, $\mathcal{M}_{\text{moduli}} =\SU(1,k)/\U(k)$ \cite{Corrado:2002wx, Louis:2015dca}, whereas in Euclidean signature $\mathcal{M}_{\text{moduli}}$ is a Wick-rotated version of the same manifold \cite{Hertog:2017owm}
\begin{equation}
\mathcal{M}_{\text{moduli}} = \frac{\SL(k+1,\mathbb{R})}{\GL(k,\mathbb{R})}\,.
\end{equation}
Euclidean supersymmetry is consistent with different Wick-rotations and to fix the above choice the following procedure was followed in \cite{Hertog:2017owm}: the moduli-space is holographically dual to the conformal manifold (space of marginal couplings) of the necklace quiver gauge theories with $k$ nodes. The marginal coouplings are $k$ complexified couplings, of which the real parts correspond to $\theta$-angles and hence get Wick-rotated with an $i$-factor. The scalars that are dual to these $\theta$ angles should be scalars that enjoy a (classical) shift symmetry (axions). The manifold
\be\label{coset2}
\mathcal{M}_{\text{moduli}} = \frac{\SU(1,k)}{\U(k)}\,,
\ee
has exactly $k$ Abelian isometries that act as shifts of $k$ real scalars. This fixes the Wick-rotation uniquely.

If we restrict to instanton solutions with spherical symmetry, the metric Ansatz is given by
\begin{equation}
\d s_5^2 = f(r)^2\d r^2 + a(r)^2\d\Omega_4^2\,,
\end{equation}
and the moduli only depend on $r$. It is well-known for such system the scalar field equations of motion reduce to geodesic equations on $\mathcal{M}_{\text{moduli}}$ and the Einstein equations of motion decouple from the scalar fields into a universal form, \cite{Breitenlohner:1987dg, Hertog:2017owm}. In the gauge $f=a^4$, $r$ is an affine parametrization of the geodesics such that
\be
G_{IJ}\dot{\phi}^I\dot{\phi}^J = c\,,
\ee
where $\dot{\phi}=\d\phi/\d r$. The metric can be determined completely in terms of this number $c$ from the ``Hamiltonian constraint'' equation
\begin{equation}\label{firstorder}
\frac{\dot{a}^2}{f^2} = \frac{c}{24}a^{-6} + \frac{a^2}{l^2} + 1\,.
\end{equation}
When $c=0$ the metric is just pure Euclidean AdS. This is due to the vanishing of the total energy momentum of the scalar fields, which is possible because of the indefinite sigma model metric. The scalar fields in that case trace out lightlike geodesics and the instantons are called \emph{extremal}.

When $c>0$ the instantons are \emph{sub-extremal} and the metric has a spike-like singularity at $r=0$ and asymptotes to AdS \cite{Bergshoeff:2005zf}
.

When $c<0$ the instantons are called \emph{super-}extremal and the geometry describes a smooth two-sided wormhole that asymptotes to AdS on both sides \cite{Gutperle:2002km, Bergshoeff:2005zf,ArkaniHamed:2007js}. Despite the smooth geometry, the scalar fields on the simplest sigma models tend to have singular scalar fields, whose singularities cancel against each other in the energy-momentum tensor. Such wormholes are considered unphysical \cite{Bergshoeff:2005zf, ArkaniHamed:2007js}. The first attempts to embed smooth solutions into AdS/CFT were described in \cite{ArkaniHamed:2007js}, whereas recently a very explicit and concrete embedding was found inside $\rm AdS_5\times S^5/\mathbb{Z}_k$ when $k>1$ \cite{Hertog:2017owm}. That observation was the inspiration for this work,  although the main goal of this paper is to  understand the extremal instantons. Some details of the non-extremal instantons are contained here as  well.

The extremal instantons can straightforwardly be extended to non-spherical solutions as follows. The affine geodesic coordinate $\tau$ equals $r$ in the gauge choice $f=a^4$ and it is obvious to check that $r$ is a harmonic function. In a different gauge, it remains of course true that the affine coordinate $\tau(r)$ equals a spherically-symmetric harmonic function on Euclidean $\rm AdS_5$
\be
\partial_r(f^{-1}a^4\partial_r \tau(r)) =0\,.
\ee
When $c=0$ the scalars do not backreact on the metric and as a consequence the replacement $\phi^i(\tau) \rightarrow \phi^i(H)$ with $H$ the most general harmonic function on Euclidean $\rm AdS_5$ still solves all equations of motion\footnote{This observation is identical to the situation with extremal Reissner-Nordstr\"om black holes, where the spherically symmetric solution can easily be extended in terms of general harmonic functions.}.  The most general harmonic function $H$ with a single center can be written in terms of the $\SO(1,5)$ invariant function:
\be
F(z, \vec{x}) = \frac{\sqrt{[(z_0-z)^2 + (\vec{x}-\vec{x}_0)^2][(z_0+ z)^2 + (\vec{x}-\vec{x}_0)^2]}}{2z}\,,\label{FFunc}
\ee
where Euclidean Poincar\'{e} coordinates are used\footnote{In these coordinates the metric is given by $\d s^2 = \frac{\ell^2}{z^2}\Bigl(\d z^2 + \d\vec{x}^2\Bigr)$.}. Now $H$ can be written as:
\be
H(z, \vec{x}) = \alpha F^{-3}\left( \left(1 - \frac{2F^2}{z_0^2}\right)\sqrt{1+\frac{F^2}{z_0^2}}\right) + \beta\,,\label{Hharm}
\ee
with $\alpha, \beta$ constants\footnote{We fix $\alpha$ and $\beta$ such that for the spherically symmetric harmonic we simply have $H=r$ in the gauge $f=a^4$, see Appendix \ref{A1}.}. The singularity of the harmonic at $z=z_0, \vec{x}=\vec{x}_0$ can be interpreted as the position of the instanton and is free to chose. So the whole of Euclidean $\rm AdS_5$ is part of the instanton moduli space. The specific choice $z_0=\ell, \vec{x}_0=0$ can be thought of as the original spherically-symmetric solution, where $H\sim \tau$. The most general solution now consists of taking arbitrary superpositions of harmonics with singularities at different places. These can be thought of as multi-centered instantons.

\subsection{The geodesic curves}
To construct the explicit geodesic curves we introduce the following $2k$ real coordinates on the moduli space:
\begin{equation}\label{coordinates}
U , a, \zeta^i ,\,\tilde{\zeta}_i\,,
\end{equation}
where $i=1\ldots k-1$. These coordinates were described in detail in \cite{Hertog:2017owm} and form the natural coordinates in a so-called solvable basis on the coset. The metric on the moduli space can be written as:
\begin{equation}\label{metric}
\d s^2=4 \d U^2-{e^{-4U}}\mathcal{N}^2+2{e^{-2U}}\sum_{i=1}^{k-1}[(\d\zeta^i)^2-(\d\tilde{\zeta}_i)^2]\,,
\end{equation}
where the one-form $\mathcal{N}$ is defined as follows
\be
\mathcal{N}\equiv da+\mathcal{Z}^M\mathbb{C}_{MN}d\mathcal{Z}^N\,,
\ee
with $\mathcal{Z}^M\equiv (\zeta^i ,\,\tilde{\zeta}_i )$ and $\mathbb{C}_{MN}$ the symplectic matrix\footnote{Explicitly we have $\mathbb{C}_{MN}=\begin{pmatrix} {\bf 0} & {\bf 1}\\ -{\bf 1} & {\bf 0} \end{pmatrix}$ in block notation.}. In contrast, the metric on the moduli space of Lorentzian AdS which (somewhat confusingly) has Euclidean signature, and can trivially be obtained from the above metric by flipping the negative signs in front of  $\mathcal{N}^2$ and $(d\tilde{\zeta}_i)^2$ .

The geodesic solutions can most easily be constructed using the exponential map:
\begin{equation}
M = M(0) \exp(2Q \tau)\,,
\end{equation}
with $Q$ an element of the Lie algebra of the coset, $\tau$ the affine coordinate and $M$ a matrix, build from the coset representative $\mathbb{L}$ (for us in solvable gauge).  The details are left for the Appendices \ref{A2},\ref{geodesiccharges},\ref{k=1}.

Geodesics through the origin have $M(0)={\bf  1}$ and are somewhat simpler. Surprisingly these exponential matrices can be completely dissected to get the explicit expressions for the separate scalar fields (\ref{coordinates}). In the appendices we have laid out the details of this construction and merely state the result here for the extremal solutions:
\begin{align}
U\,&=\,\frac{1}{2}\,\log\left[\frac{1}{\left(1-\tau p_0\right)\left(1-\tau m_0\right)}\right]\quad,\label{gensol1}\\
\zeta^i\,&=\,-\tau\,\left(\,\frac{p_i}{\sqrt{2}\left(1- \tau p_0\right)}\,+\,\frac{m_i}{\sqrt{2}\left(1-\tau m_0\right)}\,\right)\quad,\label{gensol2}\\
\tilde{\zeta}_i\,&=\,-\tau\,\left(\,\frac{p_i}{\sqrt{2}\left(1-\tau p_0\right)}\,-\,\frac{m_i}{\sqrt{2}\left(1-\tau m_0\right)}\,\right)\quad,\label{gensol3}\\
a\,&=\,-\frac{1}{\left(1-\tau p_0\right)}\,+\,\frac{1}{\left(1-\tau m_0\right)}\quad\label{gensol4},
\end{align}
 where $i=1,\dots, k-1$. There are $2k$ integration constants $p_{\alpha}, m_{\beta}$ with $\alpha, \beta=0\ldots k-1$ that obey
\be
\vec{m}\cdot \vec{p}=0.
\ee
This condition implies that the Noether charge matrix $Q$ is nilpotent. As we explain in the Appendices there are two kinds of solutions:  degree 2 ($Q^2=0$) and degree 3 ($Q^3=0$) geodesics.

For the simple case $k=1$ we show in Appendix \ref{k=1} how we reproduce the known D-instanton solutions.

\subsection{Geodesic orbits and normal forms}
The general solution is described by geodesics whose initial point  at radial infinity is different from the origin $O$ of the moduli space, and is defined by generic values of the scalar fields. These geodesics are obtained by acting on the ones originating in $O$ by means of shift-like isometry transformations:
\begin{align}
 U & \rightarrow U + U(0)\,,\nonumber\\
\tilde{\zeta} & \rightarrow \tilde{\zeta}e^{U(0)} + \tilde{\zeta}(0)\,,\nonumber\\
\zeta & \rightarrow {\zeta}e^{U(0)} + {\zeta}(0)\,,\nonumber\\
a & \rightarrow a e^{2U(0)} + \zeta\tilde{\zeta}(0)e^{U(0)} - \tilde{\zeta}\zeta(0)e^{U(0)} + a(0)\,.\label{shifstsym}
\end{align}
The above transformations are isometries in $\SL(k+1,\mathbb{R})$ that act transitively on the coset. Once the isometry is fixed that brings a general geodesics curve to a curve through the origin, there is still the freedom to play with the isotropy group that rotates the velocity vector in the origin. This allows us to bring the charge matrix $Q$ (\ref{QNoether}) to its normal form. We will do this by fixing the action of ${\rm SO}(k)$ on the Noether charge:
We  can reduce $\vec{p}$ to $\vec{p}=(p_0,0,\dots,0)$ and using the ``little group'' ${\rm SO}(k-1)$ of $\vec{p}$ we can rotate $\vec{m}$ to the form $\vec{m}=(m_0,m_1,0,\dots,0)$.
The various orbits of solutions discussed above correspond to the following choice of parameters:
\begin{align}
& Q^3=0: m_0=0,\,m_1\neq 0, \,p_0\neq 0\,,\\
& Q^2=0: p_0=0 \quad \text{or}\quad \vec{m}=\vec{0}\,.
\end{align}
The conclusion is that all lightlike geodesics can be obtained by acting with isometries of the AdS moduli space on geodesics with these simple charges.

\section{The on-shell actions}\label{sec:on-shell}
\subsection{Hodge duality}
To compute the on-shell action for the instantons one cannot use the sigma-model action (\ref{5Daction}) since it vanishes\footnote{The infinite contribution from the cosmological constant is canceled by holographic renormalisation (see for instance \cite{Bergshoeff:2005zf}).}. Instead a total derivative is needed to define action that has the correct value on the solution. For the case of a single axion-dilaton pair (\ref{axion-dilaton}) it was argued in \cite{Giddings:1987cg} that a term proportional to $\partial(\chi \exp(b\phi)\partial\chi)$  needs to be added and this total derivative provides the full on-shell action. This proposal has been shown correct, at least in the AdS/CFT context, where this prescription made the $D(-1)$ on-shell action match exactly with the on-shell action of the dual YM instantons \cite{Balasubramanian:1998de}.

A simple way to argue for that specific boundary term comes from Hodge-dualising the axion to a $3$-form potential $B_3$ with 4-form fieldstrength $H_4=\d B$. The action in terms of the Hodge dual is given by:
\be\label{B3-dilaton}
S = -\frac{1}{2\kappa_5^2} \int \sqrt{|g_5|}\Bigl(R_5 - \tfrac{1}{2}(\partial\phi)^2 - \tfrac{1}{2 (4!)}\e^{-b\phi} H_{\mu_1\ldots \mu_4}H^{\mu_1\ldots \mu_4} - \Lambda \Bigr)\,.
\ee
Note that there is no flipped sign of the kinetic term here, neither in Euclidean nor in Lorentzian signature. If the path integral is considered in terms of this Hodge dual field configuration it is easy to argue that performing Hodge duality by adding Lagrange multipliers provides the action (\ref{5Daction}) plus the required total derivative \cite{Bergshoeff:2005zf}. In other words, using the Hodge dual action directly provides the correct answer for the on-shell action.

We now apply the same logic to our more sophisticated sigma model: we will Hodge dualise all axions and use the resulting action (without boundary terms) to compute the on-shell action. The proper way to Hodge dualise proceeds via adding Lagrange multipliers that are 3-form potentials $C_3$. To  Hodge dualise we need to make the shift symmetries manifest by using $\tilde{a}\equiv a-\zeta^i\,\tilde{\zeta}_i$ instead of $a$:
\begin{equation}
\mathcal{N} = \d \tilde{a} + 2 \zeta^i\d \tilde{\zeta}_i\,.
\end{equation}
Now $\tilde{a}$ and the $\tilde{\zeta}_i$ appear explicitly shift-symmetric and can be dualised to $3$-forms. From the EOM, the conserved $4$-form field strengths are
\begin{align}
& H_0 = \star e^{-4U}\mathcal{N}\,,\label{Hodge1}\\
& H_i = \star\Bigl( e^{-4U}\mathcal{N}\zeta^i + e^{-2U} \d \tilde{\zeta}_i\Bigr)\,,\label{Hodge2}
\end{align}
with $i=1,\ldots, k-1$. These are the Hodge duals to the magnetic 1-form fieldstrengths $F_1 = \d \tilde{a}$ and $F_i = \d \tilde{\zeta}_i$.
We now use Legendre transformations in order to dualise the action. The reason for presenting these details is that this procedure generates the required boundary term that leads to a finite instanton action.

We start from the sigma model action (\ref{5Daction}). To perform the Legendre transform one replaces $ d\tilde{a}\,\rightarrow\,\,\,F_0$ and $\d\tilde{\zeta}_i\,\rightarrow\,\,\,F_i$
and regards  $F_0, F_i$ as auxiliary 1-form fields. Next we add Lagrange multiplier terms so to obtain the following action:
\begin{align}
S'=&S_{{\rm grav}}+\frac{1}{2\kappa_5^2}\,\int 2\,\d U\wedge \star \d U+e^{-2U}\, \d\zeta^i\wedge \star\d\zeta^i-e^{-2U}\,F_i\wedge \star F_i \nonumber\\
&-\frac{e^{-4U}}{2}\,(F_0+2\,\zeta^i\,F_i)\wedge \star(F_0+2\,\zeta^i\,F_i)-2 H_i\wedge (\d\tilde{\zeta}_i-F_i)-H_0\wedge\,(\d\tilde{a}-F_0)\,,\label{Sprime}
\end{align}
where sum over repeated index $i$ is understood and form-notation was used\footnote{$\omega_{(p)}\wedge \star\omega_{(p)}=(-1)^{(D-p)p}\,\frac{1}{p!}\,\omega_{\mu_1\dots \mu_p}\omega^{\mu_1\dots \mu_p}\,\,,\,\,\,\,\star\star\omega_{(p)}=(-1)^{(D-p)p}\,\omega_{(p)}$, where in our case $D=5$. }.
Extremizing (\ref{Sprime}) with respect to $H_0, H_i$ we find $d\tilde{\zeta}_i=F_i,\,d\tilde{a}=F_0$ and we are back to the original Lagrangian (\ref{5Daction}). Extremizing, on the other hand, with respect to $\tilde{a},\,\tilde{\zeta}_i$ we find:
\begin{equation}
\d H_i=\d H_0=0\,\,\Rightarrow\,\,\,\,H_i=\d C_i\,,\,\,H_0=\d C_0\,.
\end{equation}
Finally extremizing with respect to $F_i$ and $F_0$, we end up with the dual action \emph{modulo} boundary terms from $H_0\wedge\,\d\tilde{a},\,2 H_i\wedge \d\tilde{\zeta}_i$:
\begin{equation}\label{Stotal}
S'=S^{({\rm dual})}+ S^{({\rm bdry})}\,,
\end{equation}
where\footnote{If $\omega_{(p)}$ is a $p$-form, we define $\omega_{(p)}^2=\omega_{(p)\,\mu_1\dots\mu_p}\,\omega_{(p)}{}^{\mu_1\dots\mu_p}$.}
\begin{align}\label{dualaction}
S^{({\rm dual})} =&   -\tfrac{1}{2\kappa_5^2}\int\sqrt{g}\Bigl(R - 2(\partial U)^2 - \tfrac{1}{2}\tfrac{1}{4!}e^{4U}H_0^2 - e^{-2U}\Bigl(\sum_i (\partial \zeta^i)^2 -  \tfrac{1}{4!}e^{4U}G_i^2\Bigr)\Bigr)\,,\\
S^{({\rm bdry})}=&-\tfrac{1}{2\kappa_5^2}\int d\mathcal{L}_B\,,\,\,\,\,\,\,\,\,\,\,\,\,\,\,\,\,\mathcal{L}_B=2\,H_i\,\tilde{\zeta}_i+H_0\, \tilde{a}\,,\label{dualboundary}
\end{align}
where we defined the combination $G_i \equiv H_i - \zeta^i H_0$.

\subsection{On-shell action as a boundary integral}
In the previous section we have dualized the axions $\tilde{a},\,\tilde{\zeta}_i$ into $3$-forms. From (\ref{Stotal}) and the vanishing of on-shell sigma model action ($S'$), we infer that
\be
S^{({\rm dual})} = - S^{({\rm bdry})}\,.
\ee
Hence if the dual action is considered as fundamental, because it has no unusual signs of kinetic terms, we deduce
\be
S_{{\rm on-shell}} =  - S^{({\rm bdry})}\,,
\ee
which means we simply have to evaluate a boundary term and there is no need to integrate. Since the on-shell action has also an imaginary component we will from here on write
\be
S^{({\rm real})}_{{\rm on-shell}}=  - S^{({\rm bdry})}\,.
\ee

To evaluate the above boundary term, we use the expressions of $H_i,\,H_0$ in terms of the Noether currents associated with the shifts in $\tilde{\zeta}_i,\,\tilde{a}$ to arrive at:
\be\label{onshellasboundary}
S^{({\rm bdry})}_{{\rm solution}}=-\tfrac{1}{2\kappa_5^2}\int_{\partial {\rm EAdS}_5}\mathcal{L}_B\,,
\ee
with
\begin{equation}
\mathcal{L}_B=e^{-4U}\,(a+\tilde{\zeta}_i\,\zeta^i)\,*(da+\zeta^i \,d\tilde{\zeta}_i-\tilde{\zeta}_i\,d \zeta^i )+2\,e^{-2U}\,\tilde{\zeta}_i\,*d\tilde{\zeta}_i\,.
\end{equation}
Hence we find:
\begin{equation}
S^{({\rm bdry})}_{{\rm solution}}=-\frac{Vol(S^4)}{2\,\kappa^2_5}\,\Bigl[\Pi(\tau=\infty)-\Pi(\tau=0)\Bigr]\,,\label{boundintegral}
\end{equation}
where
\begin{equation}
\Pi(\tau)=e^{-4U}\,(a+\tilde{\zeta}_i\,\zeta^i)\,(\dot{a}+\zeta^i \,\dot{\tilde{\zeta}}_i-\tilde{\zeta}_i\,\dot{\zeta}^i )+2\,e^{-2U}\,\tilde{\zeta}_i\,\dot{\tilde{\zeta}}_i\,.
\end{equation}
Note that, in our choice of parametrization of the geodesic, radial infinity (``the UV'') corresponds to $\tau=0$, which is where the dual boundary theory lives, whereas the ``IR'' is towards $\tau=\infty$.

Regular solutions require $p_0$ and $m_0$ to have the same sign and by carefully evaluating the expression (\ref{boundintegral}), one finds the following, manifestly positive action:
\begin{equation}
S^{({\rm real})}_{{\rm on-shell}}=\frac{\rm  Vol(S^4)}{2\,\kappa^2_5}\,\left[|(m_0+p_0)|\left(1 + \tfrac{1}{2}\Bigr[\sum_{i=1}^{k-1}\frac{m_i^2}{m_0^2}+\sum_{i=1}^{k-1}\frac{p_i^2}{p_0^2}\Bigl]\right)\right]
\,.\label{boundintegral1}
\end{equation}
Supersymmetric solutions have all $p_\alpha=(p_0,\,p_i)$ or all $m_\alpha=(m_0,\,m_i)$ equal to zero. If we consider the case all $p_i$ to vanish, the on-shell action becomes:
\begin{equation}
S^{({\rm real})}_{{\rm on-shell}}=\frac{\rm  Vol(S^4)}{2\,\kappa^2_5}\,\frac{1}{|m_0|}\left(m_0^2+ \frac{1}{2}\sum_{i=1}^{k-1}m_i^2\right)\,.\label{boundintegralBPS}
\end{equation}

\subsection{Imaginary part of the action and charge quantisation}

Now we turn to the imaginary part of the action, based on the appendix of \cite{Bergshoeff:2005zf}. The path integral quantisation  entails that for every scalar that is shift symmetric (and which will be dualised) one simply adds its boundary value times the axion charge. The axion charges are easily computed:
\begin{align}
& q_0 =\text{Vol}(S^4)^{-1}\int_{S^4} H_0 =   e^{-2U(0)}(m_0 - p_0) \,,\\
& q_i =\text{Vol}(S^4)^{-1}\int_{S^4} H_i =  \frac{e^{-U(0)}}{\sqrt{2}}(m_i -p_i) + e^{-2U(0)}(m_0 - p_0)\zeta^i(0)\,.
\end{align}
The boundary in this context means the physical boundary (i.e. UV) of the Euclidean AdS space and corresponds to $\tau=0$.   So we have
\begin{equation}
S_{\text{on-shell}}^{\rm imaginary} =  \frac{\rm Vol(S^4)}{2\,\kappa^2_5}\Bigl(i \tilde{a}(0)\,q_0 + 2i\sum_j\tilde{\zeta}_j(0)\,q_j\Bigr)\,,
\end{equation}
or, written differently using previous notation,
\begin{equation}
S_{\text{on-shell}}^{\rm imaginary} =  i\,\frac{\rm Vol(S^4)}{2\,\kappa^2_5}\Pi(0)\,.
\end{equation}
As opposed to the real part of the on-shell action, the imaginary part is not invariant under shifts of the axion. The shift invariance of the real part is due to the subtraction $\Pi(\infty)-\Pi(0)$. For geodesics through the origin this contribution is zero. The field theory dual interpretation of the imaginary part is  the well known $i\theta \text{Tr}{F\wedge F}$ contribution and the dual $\theta$'s are nothing but the boundary values of the axions. So if they all vanish, as is the case of geodesics through the origin, the imaginary action vanishes.

In $\rm AdS_5\times S^5$ there was a match between the real part of the on-shell action of the supergravity and the dual gauge instanton. But also, following the above procedure, a match between the imaginary pieces is achieved (see for instance \cite{Bergshoeff:2005zf}).

The axion charges should be quantised and the exact quantisation condition depends on the fundamental domain of the moduli space. In other words, it depends on the identification of the axion to itself:
\be \label{shiftaxion}
\tilde{a} = \tilde{a} + L_0\,,\quad \tilde{\zeta}_i= \tilde{\zeta}_i + L_i\,,
\ee
where $L_0$ and $L_i$ are the lengths of the axion-circles. What these constants $L$ should be depends on the microscopic theory. So either one starts off with the 10D string theory and analyses the dimensional reduction over $S^5/\mathbb{Z}_k$ to identify the 10D origin of the axions, or one uses the detailed map between the moduli and the dual gauge couplings of the quiver. We leave this for future investigation  and for now just state the quantisation in terms of the circle lengths. For instance, following the recent discussion in \cite{Alonso:2017avz}, we simply use that the boundary action (\ref{dualboundary}) should shift as $2\pi n$, with $n$ integer, under the shifts of the axions (\ref{shiftaxion}). This implies the following quantisation rules
\begin{align}
& q_0 = e^{-2U(0)}(m_0 - p_0) = n_0\frac{\kappa_5^2}{\rm Vol(S^4)}\frac{2\pi}{L_0}\,,\nonumber\\
& q_i =
\frac{e^{-U(0)}}{\sqrt{2}}(m_i -p_i) + e^{-2U(0)}(m_0 - p_0)\zeta^i(0) = n_i\frac{\kappa_5^2}{2 \rm Vol(S^4)}\frac{2\pi}{L_i}\,,
\end{align}
with $n_0, n_i \in \mathbb{Z}$.

\section{Supersymmetry}\label{sec:susy}

We now consider the supersymmetry properties of the extremal solutions in (\ref{gensol1}-\ref{gensol4}).
  To this end  we Wick-rotate the solutions to complex solutions of Lorentzian $\mathcal{N}=4,\,D=5$ gauged theory, for which we know the supersymmetry transformation rules \cite{Schon:2006kz,Louis:2015dca}.
The Wick-rotated extremal solutions are trivially obtained from (\ref{gensol1}-\ref{gensol4}) by multipling $a$ and $\tilde{\zeta}'s$ with an $i$.  Now $\tau$ is a harmonic function on $\rm AdS_5$. Those solutions,  in spite of being complex, solve the geodesic equations on ${\rm SU}(1,k)/{\rm U}(k)$ and thus the field equations of the Lorentzian $\mathcal{N}=4,\,D=5$ gauged supergravity.

We now  review the relevant features of this theory and the embedding of the moduli space ${\rm SU}(1,k)/{\rm U}(k)$ inside the corresponding scalar manifold.

\subsection{Half-maximal gauged supergravity}

Half-maximal supergravity in $4+1$ dimensions has a scalar manifold of  the following general form:
\begin{equation}\label{coset3}
\mathcal{M}_{scal}={\rm SO(1,1)\times \frac{SO(5,n)}{SO(5)\times SO(n)}}\,.
\end{equation}
We denote by $\Sigma$ the scalar parametrizing the ${\rm SO(1,1)}$-factor and by $\mathcal{V}_M{}^N$ the coset representative of the latter factor in the fundamental representation of ${\rm SO(5,n)}$ and thus satisfies the condition:
\begin{equation}
\mathcal{V}_M{}^P\mathcal{V}_N{}^Q\,\eta_{PQ}=\mathcal{V}_M{}^m\mathcal{V}_N{}^m-\mathcal{V}_M{}^a\mathcal{V}_N{}^a=\eta_{MN}\,,
\end{equation}
where
\begin{equation}
\eta_{MN}\equiv {\rm diag}(+1,+1,+1,+1,+1,-1,\dots,-1)\,\,,
\end{equation}
and we have written $P,\,Q=(m,a)$, $m=1,\dots,5\,,\,\,\,a=6,\dots, n+5$.

We now closely follow \cite{Louis:2015dca}. The most general gauging of the theory is defined by an embedding tensor which consists of the ${\rm SO}(5,n)$-tensors, $\xi_M,\,\xi_{MN}$ and $f_{MNP}$, satisfying  suitable linear and quadratic constraints \cite{Schon:2006kz}. For the case of interest we can restrict to $\xi_M=0$ and then the remaining tensors satisfy
\begin{align}
& \xi_{MN}=-\xi_{NM}\,,\qquad \xi^{MQ}\,f_{QNP}=0\,,\nonumber\\
& f_{MNP}=f_{[MNP]}\,,\quad f_{RM[N} f_{PQ]}{}^R=0\,.
\end{align}
The gauge generators $T_0,\,T_M$ are defined, in the fundamental representation of ${\rm SO}(5,n)$, as
\begin{equation}
(T_0)_N{}^P=\xi_N{}^P\,\,,\,\,\,\,(T_M)_N{}^P=f_{MN}{}^P\,,
\end{equation}
and close the algebra:
\begin{equation}
[T_0,\,T_M]=0\,\,,\,\,\,\,[T_M,\,T_N]=-f_{MN}{}^P\,T_P\,.
\end{equation}
We further specialize the two tensors to have the following non-zero entries:
\begin{align}
\xi^{12}=\xi^{67}=\dots=\xi^{2\ell+4,2\ell+5}\,,\,\,\ell=1,\dots,k\,,\nonumber\\
f_{345}\,\,,\,\,\,\,f_{a'b'c'}\,\,,\,\,\,\,a',b',c'=2k+6,\dots, n+5\,.\label{xiform}
\end{align}
The number $k$ corresponds to the order of the orbifold group $\mathbb{Z}_k$.  We leave the further specifications of the gauge group for later, after we have introduced the supersymmetry transformations.

To write the fermion transformation rules it is also useful to introduce the ${\rm SO}(5)$ gamma-matrices $(\Gamma^m)_i{}^j$, $i,j=1,\dots,4$, whose explicit form can be found in Appendix \ref{Cliff}.
In particular we define
\begin{equation}
\mathcal{V}_M{}^{ij}=\mathcal{V}_M{}^{m}\,(\Gamma_m)^{ij}\,,
\end{equation}
and the anti-symmetric matrix
\be
\Omega_{ij}=(\Gamma^4\Gamma^2)_{ij}\,,
\ee
whose details are also laid out in Appendix \ref{Cliff}.

The supersymmetry transformations for the four gravitini $\psi^i_{\mu}$, the four spin-$1/2$ fermions $\chi^i$ and the gaugini $\lambda^i_a$ are given by \cite{Schon:2006kz,Louis:2015dca}
\begin{align}
\delta\psi_{\mu i}&=D_{\mu}\epsilon_i+\frac{\textit{i}}{\sqrt{6}}\Omega_{ij}A_1^{jk}\Gamma_{\mu}\epsilon_k+\dots\label{deltapsi}\quad,\\
\delta\chi_{i}&=-\frac{\sqrt{3}}{2}\,i\,\Sigma^{-1}\,D_\mu \Sigma\,\Gamma^\mu\epsilon_i+\sqrt{2}\,A_2^{kj}\epsilon_k+\dots\label{deltachi}\quad,\\
\delta\lambda_{i}^a&=i\,\Omega^{jk}\,\mathcal{V}^{-1}{}_M{}^{a}D_\mu \mathcal{V}_{ij}{}^M\,\Gamma^\mu\epsilon_k+\sqrt{2}A_2^{a\,kj}\epsilon_k+\dots\label{deltalambda}\quad,
\end{align}
where $\epsilon_j$ are the usual four supersymmetry parameters and the $\ldots$ indicate terms involving the vector field strengths. Here $\Gamma^\mu$ denote the space-time gamma-matrices (not to be mistaken with the ${\rm SO(5)}$ matrices $\Gamma^m$). The fermion shift matrices $A_1^{jk}$, $A_2^{kj}$ and $A_2^{a\,kj}$ entering these transformations are defined as
\begin{align}
A_1^{ij}&=-\frac{1}{\sqrt{3}} \Sigma^{2}\Omega_{kl}\mathcal{V}^{ik}_M\mathcal{V}^{jl}_N\xi^{MN}-\frac{4}{3\sqrt{6}}\Sigma^{-1} \mathcal{V}^{ik}_M \mathcal{V}^{jl}_N \mathcal{V}_{kl}^P {f^{MN}}_P\,,\label{shift1}\\
A_2^{ij}&= \frac{1}{\sqrt{3}}\Sigma^{2}\Omega_{kl}\mathcal{V}^{ik}_M\mathcal{V}^{jl}_N\xi^{MN} -\frac{2}{3\sqrt6}\Sigma^{-1} \mathcal{V}^{ik}_M \mathcal{V}^{jl}_N \mathcal{V}_{kl}^P {f^{MN}}_P + \frac{3}{2\sqrt6}\Sigma^{-1}\mathcal{V}^{ij}_M\xi^M \,,\label{shift2}\\
A_2^{a\,ij}&=-\frac{1}{\sqrt{2}}\Sigma^{2}\mathcal{V}^{a}_M\mathcal{V}^{ij}_N\xi^{MN} + \frac{1}{\sqrt{2}}\Sigma^{-1}\Omega_{kl}\mathcal{V}^{a}_M \mathcal{V}^{ik}_N \mathcal{V}^{jl}_P f^{MNP}-\frac{\sqrt{2}}{8}\Sigma^{-1}\mathcal{V}^{a}_M\xi^M\Omega^{ij}\,.\label{shift3}
\end{align}
In terms of these tensors, the scalar potential is then given by
\begin{equation}
\frac{1}{4}\Omega^{ij}V\,=\,\Omega_{kl}\left(A_2^{a\,ik}A_2^{a\,jl}+A_2^{ik}A_2^{jl}-A_1^{ik}A_1^{jl}\right)\,\,\,.
\end{equation}
The vanishing of the supersymmetry transformations in (\ref{deltapsi}-\ref{deltalambda}) in the $\rm AdS_5$ background, where all supercharges are unbroken, entails
\begin{align}
&\langle A_2^{ij} \rangle\,=\,\langle A_2^{a\,ij} \rangle\, =\,0\,,\label{cA1}\\
&\langle A_1^{ij} A_{1\,kj} \rangle\,=\,\frac{1}{4} |\mu|^2 \delta^i_k \,.\label{cA2}
\end{align}
These constraints were solved in \cite{Louis:2015dca} where general conditions on the gauging parameters, compatible with the existence of the $\rm AdS_5$ vacuum, were defined. These conditions are solved by the choice \eqref{xiform}, with
\begin{equation}
\xi_{12}=\frac{\sqrt{2}}{\Sigma^3}\,f_{345}\,,
\end{equation}
where in the vacuum we can fix $\Sigma=1$. We choose the gauge group $\mathcal{G}$ to have the following general form
\begin{equation}\label{gauge}
\mathcal{G}={\rm U(1)\times SU(2)\times H_c}\,,
\end{equation}
where we could take for instance ${\rm H_c=SU(2)}$ by choosing $n=2k+3$ and $f_{a'b'c'}=f\,\epsilon_{a'b'c'}$, although the particular choice of ${\rm H_c}$ will not be relevant to our discussion\footnote{To be more concrete,
one can show that the orbifold compactification leads to the gauging $SU(2)\times SU(2)\times U(1)$ when $k=2$ and to $SU(2)\times U(1)$ when $k>2$ \cite{Corrado:2002wx}. }.

\subsection{Instanton Killing spinor equations}

We now compute the supersymmetry variations of the fermion fields on the instanton backgrounds. Since only moduli scalars are switched on, the fermion shift tensors still satisfy \eqref{cA1}-\eqref{cA2}. In particular the $A_2$-tensors vanish. The only new terms in the fermion supersymmetry transformation rules, with respect to the vacuum case, are those involving the space-time derivatives of the scalar fields and of the supersymmetry parameters. The dependences are with respect to $\tau$. As indicated before, from here onwards,  the variable $\tau$ is allowed to be any harmonic on $\rm AdS_5$ and so is not necessarily the radially symmetric harmonic.

Supersymmetry requires that supersymmetry parameters $\epsilon_i$ exist such that
\begin{align}
D_{\mu}\epsilon_i+\frac{\textit{i}}{\sqrt{6}}\Omega_{ij}A_1^{jk}\Gamma_{\mu}\epsilon_k&=0\,,\label{Kill1}\\
\Omega^{jk}\,\mathcal{V}^{-1}{}_M{}^{a}D_\mu \mathcal{V}_{ij}{}^M\,\Gamma^\mu\epsilon_k&=0\,.\label{Kill2}
\end{align}
We seek for a solution of the above equations of the form $\epsilon^i={\bf g}(\tau)_i{}^j\,\mathring{\epsilon}^{\,j}$, where $\mathring{\epsilon}^{\,i}$ are the four Killing spinors of the vacuum.
We have fixed $\Sigma=1$, so that the variation of the dilatinos does not imply any new condition.

Let us first solve equation \eqref{Kill2}.
The matrix $\mathcal{V}$ entering its right-hand-side is evaluated on the solution, which depends on the space-time coordinates only through the harmonic function $\tau$. Therefore, denoting by $\Gamma$ the following space-time dependent matrix
\begin{equation}
\Gamma\equiv \Gamma^\mu\,\partial_\mu \tau\,,
\end{equation}
equation \eqref{Kill2} can be recast in the form
\begin{align}
\Omega^{jk}\,\mathcal{V}^{-1}{}_M{}^{a}\dot{\mathcal{V}}_{ij}{}^M\,\Gamma\,\epsilon_k&=0\,.\label{Kill3}
\end{align}
Note that the composite connection $\mathcal{Q}_\mu$ of the scalar manifold does not contribute to the covariant derivative in \eqref{Kill2} since $\mathcal{Q}_{\mu\, i}{}^a=0$. Equation \eqref{Kill3} further simplifies if we notice that $\Gamma$ is a non-singular matrix, so that the condition can be written as follows:
\begin{align}
N^a{}_i{}^k\,\epsilon_k\equiv\Omega^{jk}\,\mathcal{V}^{-1}{}_M{}^{a}\dot{\mathcal{V}}_{ij}{}^M\,\epsilon_k&=0\,.\label{Kill4}
\end{align}
This equation implies that the matrices ${\bf N}^a=(N^a{}_i{}^k)$ must be singular and have a common null vector. The determinants of these matrices are found to be:
\begin{align}
\det({\bf N}^{2i+4})&=\det({\bf N}^{2i+5})\propto\,\frac{m_i^2 p_i^2}{(1-\tau\,p_0)^2(1-\tau\,m_0)^2}\,\,,\,\,\,\,i=1,\dots, k-1\,,\nonumber\\
\det({\bf N}^{a})&=0\,\,,\,\,\,a=5+2k,\dots, n+5\,.
\end{align}
The vanishing of the above determinants implies that $m_i p_i=0$ for each $i$ ($>0$). From the nilpotency condition $\vec{p}\cdot \vec{m}=0$, it further follows that $m_0 p_0=0$. This condition then also implies the vanishing of $\det({\bf N}^{4+2k})$ and $\det({\bf N}^{5+2k})$ such that:
\begin{equation}
\det({\bf N}^{a})=0\,\,\,\Rightarrow\,\,\,\,m_{\alpha} p_{\alpha}=0\,\,,\,\,\,\forall \alpha=0,\dots , k-1\,.
\end{equation}
Note that one can always find representatives of the nilpotent orbits of $Q$ for which $m_{\alpha} p_{\alpha}=0$, $\forall \alpha$. We observe that the matrices ${\bf N}^{a}$ evaluated on these extremal representatives, are all nilpotent and, as we illustrate below, have a definite grading with respect to the matrix $D$:
\begin{align}
{\bf D}=\left(
          \begin{array}{cccc}
            0 & -\frac{1}{2} & 0 & 0 \\
            -\frac{1}{2} & 0 & 0 & 0 \\
            0 & 0 & 0 & \frac{1}{2} \\
            0 & 0 & \frac{1}{2} & 0 \\
          \end{array}
        \right)=-\frac{i}{2} \Gamma_1 \Gamma_2\quad,
\end{align}
which allows us to better understand the supersymmetry properties of the solution.

\subsection{$Q^3=0$ orbit}

For the orbit with $Q^3=0$ and $Q^2\neq 0$ both vectors $\vec{m}$ and $\vec{p}$ are non-vanishing. Then the ${\bf N}^{a}$ matrices are all nilpotent but without  common null vector. For instance, taking $p_1=0,\,m_2=0$ but $m_1\neq 0$ and $p_2\neq 0$, we find:
\begin{align}
{\bf N}^{8}&\propto\,{\bf N}_+\equiv\left(
\begin{array}{cccc}
 0 & 0 & -1 & 1 \\
 0 & 0 & 1 & -1 \\
 -1 & -1 & 0 & 0 \\
 -1 & -1 & 0 & 0
\end{array}
\right)=-\left(\Gamma_1+i \Gamma_2\right)\,\,,\label{N6}\\
{\bf N}^{6}&\propto {\bf N}_-\equiv\left(
\begin{array}{cccc}
 0 & 0 & -1 & -1 \\
 0 & 0 & -1 & -1 \\
 -1 & 1 & 0 & 0 \\
 1 & -1 & 0 & 0
\end{array}
\right)=-\left(\Gamma_1-i \Gamma_2\right)=\left(\Omega\right)^{T} {\bf N}^{8} \Omega\label{N8}\,.
\end{align}
One can verify that the following commutation relations hold
 \begin{align}
\left[{\bf D},\,{\bf N}_\pm\right]\,&=\pm {\bf N}_\pm\,.
\end{align}
The above nilpotent matrices annihilate no common non-vanishing vector and the corresponding solutions are not supersymmetric. Since the isometries of the scalar manifold commute with supersymmetry we deduce that this must be true for the whole orbit. Nonetheless we demonstrate this explicitly, for the sake of completeness, by analyzing the grading-structure of the ${\bf N}$-matrices in section \ref{sec:grading}.

\subsection{$Q^2=0$ orbit}
As far as the  $Q^2=0$ orbit is concerned, in which either $\vec{p}=0$ or $\vec{m}=0$, all the shift matrices have the same grading and thus annihilate the same 2-parameter spinor. More details of this can be found in the next subsection, but for now it suffices to know that  equation \eqref{Kill2} can be solved completely since their is a common kernel for the shift matrices. To show that the solutions are indeed $1/2$-BPS, we have to solve the gravitino Killing spinor equation \eqref{Kill1}.

The right-hand-side, in terms of $\epsilon_i={\bf g}(\tau)_i{}^j\,\mathring{\epsilon}_{j}$, reads:
\begin{align}
&D_{\mu}\epsilon_i+\frac{\textit{i}}{\sqrt{6}}\Omega_{ij}A_1^{jk}\Gamma_{\mu}\epsilon_k=\nonumber \\
&\left((\partial_\mu {\bf g} {\bf g}^{-1})_i{}^j+\mathcal{Q}_\mu{}_i{}^j\right)\,\epsilon_j+{\bf g}_i{}^\ell\,\left[\delta_\ell^j\left(\partial_\mu +\frac{1}{4}\,\omega_{ab,\,\mu}\,\Gamma^{ab}\right)+\frac{\textit{i}}{\sqrt{6}}\Omega_{\ell k}A_1^{kj}\Gamma_{\mu}\right]\,\mathring{\epsilon}_{j}=0\,,\label{Kill12}
\end{align}
provided we choose ${\bf g}_i{}^j$ so that it commutes with $A_{1\,i}{}^j$.
In the above equation $\mathcal{Q}_\mu{}_i{}^j$ is the pull-back on the background of the $R$-symmetry connection on the scalar manifold.
The terms in square brackets vanish being the gravitino Killing spinor equations for the vacuum.
We are left with the following condition for the matrix ${\bf g}$:
\begin{align}
\left(\partial_\mu {\bf g} {\bf g}^{-1})_i{}^j+\mathcal{Q}_\mu{}_i{}^j\right)\,\epsilon_j=0\,\,\Leftrightarrow\,\,\,\,\partial_\mu\tau\left((\dot{\bf g} {\bf g}^{-1})_i{}^j\, +\mathcal{Q}{}_i{}^j\right)\,\epsilon_j=0\,,
\end{align}
where we have used the fact that both ${\bf g}_i{}^j$ and the scalar fields only depend on space-time through the harmonic function $\tau$, defining $\mathcal{Q}{}_i{}^j$ so that: $\mathcal{Q}_\mu{}_i{}^j=\partial_\mu\tau\,\mathcal{Q}{}_i{}^j$.
We find that $\mathcal{Q}{}_i{}^j$ is proportional to the matrix ${\bf D}$ defined above:
\begin{equation}
\mathcal{Q}{}_i{}^j=-2\frac{(m_0-p_0)}{(1-\tau\,m_0)(1-\tau\,p_0)}\,{\bf D}_i{}^j\,.
\end{equation}
Notice that the above matrix is non-compact and thus it is not in ${\rm USp}(4)$ since the Wick-rotated solution on which we compute the connection is complex\footnote{Indeed, in the Euclidean version of the gauged supergravity, the R-symmetry group is non-compact.}.
We can find a matrix ${\bf g}_i{}^j$ satisfying the equation:
\begin{equation}
(\dot{\bf g} {\bf g}^{-1})_i{}^j\, +\mathcal{Q}{}_i{}^j=0\,.
\end{equation}
It suffices to take
\begin{equation}{\bf g}=e^{h(\tau)\,{\bf D}}\,,\label{gD}\end{equation}
where $h(\tau)$ is:
\begin{equation}
h(\tau)=2\,\int\,\frac{(m_0-p_0)}{(1-\tau\,m_0)(1-\tau\,p_0)}d\tau\,.
\end{equation}
In deriving Equation \eqref{Kill12} we also used the property that $A_{1\,i}{}^j$ commutes with ${\bf g}_i{}^j$, which follows from Eq. (\ref{gD}) and the property that, on our background, $$A_{1\,i}{}^j\propto {\bf D}_i{}^j\,.$$
We conclude that the gravitino Killing spinor equations \eqref{Kill1} are solved by suitably choosing the space-time dependence of the two solutions of the \eqref{Kill2} equations. This implies that the $Q^2=0$ orbit consists of $1/2$-BPS solutions.

\subsection{Further details of the Killing spinor analysis}\label{sec:grading}
In this section we  study the structure of the ${\bf N}^{a}$ in some more detail.

The shift matrices ${\bf N}^{a}$ in the extremal solution are al proportional to the nilpotent matrices ${\bf N}_\pm$ (defined in \eqref{N6}-\eqref{N8}) with coefficients depending on the charges. To show this it is useful to define the following matrices:
\begin{align}
\mathcal{N}_{\pm}^{i}&={\bf N}^{(2i+4)}\mp i{\bf N}^{(2i+5)}\,,\quad i=1,\dots,k-1\,,\nonumber \\
\mathcal{N}_{\pm}^{0}&={\bf N}^{(2k+4)}\mp i{\bf N}^{(2k+5)}\,,\label{Napm}
\end{align}
and the functions:
\begin{equation}
 \xi_{+}=\,\frac{\sqrt{(1-\tau\,m_0)(1-\tau\,p_0)}}{(1-\tau\,p_0)^2}\,,\qquad \xi_{-}=\,\frac{\sqrt{(1-\tau\,m_0)(1-\tau\,p_0)}}{(1-\tau\,m_0)^2}\,.
\end{equation}
Then the explicit form of the shift matrices ${\bf N}^{a}$ is:
\begin{align}
\mathcal{N}_{+}^{j}&=p_j \xi_{+} {\bf N}_+\,,\\
\mathcal{N}_{-}^{j}&=m_j \xi_{-} {\bf N}_-\,,
\end{align}
if $j=1,\dots,k-1$ and
\begin{align}
\mathcal{N}_{+}^{0}&=\left[p_0\sqrt{\frac{(1-\tau\,p_0)}{(1-\tau\,m_0)}} \xi_{+}+\frac{\tau^2\left(m_0-p_0\right)}{2(1-\tau\,m_0)(1-\tau\,p_0)}\vec{p}\cdot\vec{m} \right]{\bf N}_+\,\,\,,\label{Np+}\\
\mathcal{N}_{-}^{0}&=\left[m_0\sqrt{\frac{(1-\tau\,m_0)}{(1-\tau\,p_0)}} \xi_{-}-\frac{\tau^2\left(m_0-p_0\right)}{2(1-\tau\,m_0)(1-\tau\,p_0)}\vec{p}\cdot\vec{m} \right]{\bf N}_-\,\,\,.\label{Np-}
\end{align}
By definition the matrices $\mathcal{N}_{\pm}^{j}$ are nilpotent with grading
 \begin{align}
\left[{\bf D},\mathcal{N}_{\pm}^{\alpha}\right]\,&=\pm \,\mathcal{N}_{\pm}^{\alpha}\,\,,\,\,\,\,\alpha=0,\dots, k-1\,.
\end{align}

For non-extremal solutions the  ${\bf N}^a$ are expressed through \eqref{Napm} as non-nilpotent combinations of these matrices. Once the nilpotency condition on $Q$ is imposed in (\ref{Np+}) and (\ref{Np-}), and in particular $m_{\alpha} p_{\alpha}=0$, $\forall \alpha$, however, all $\mathcal{N}_{+}^{i}$ and $\mathcal{N}_{-}^{i}$ matrices are proportional to $p_i \xi_{+} {\bf N}_+$ and $m_i \xi_{-} {\bf N}_-$, respectively and, as it follows from \eqref{Napm}, the ${\bf N}^a$ themselves become nilpotent.

From the above equations it is clear that if the solution is in the $Q^3=0$ orbit, the matrices ${\bf N}^{a}$ are proportional, for different values of $a$, to matrices  $\mathcal{N}_{\pm}^{i}$ with different gradings and thus they can not have a common non-vanishing eigenvector with zero eigenvalue. In the example given earlier, if $p_1=0$ and $m_2=0$, but $m_1\neq 0$ and $p_2\neq 0$, we see that ${\bf N}^6=i\,{\bf N}^7\propto {\bf N}_-$ and  ${\bf N}^8=i\,{\bf N}^9\propto {\bf N}_+$.

Finally, we comment on the geometrical meaning of the matrix ${\bf D}$.
The gauge group breaks the ${\rm USp}(4)$ R-symmetry group into ${\rm U(1)\times SU(2)}$, which commutes with the generators of ${\rm SU}(1,k)$ inside ${\rm SO}(5,n)$, since the moduli are singlets with respect to it. From equation (\ref{embedding}) we see that the generators of ${\rm SU}(1,k)$ are embedded in the fundamental representation
of ${\rm SO}(5,n)$ as matrices with non-trivial entries in the rows and columns labelled by the values  $\tilde{m}=1,2$, $\tilde{a}=1,\dots, 2k$ of the indices $m=1,\dots, 5$ and $a=1,\dots,n$.
The ${\rm U}(1)$ gauge generator $J_{0}$ in the same representation of ${\rm SO}(5,n)$ reads \cite{Louis:2015dca}:
\begin{equation}
J_0=\,{\rm diag}(\boldsymbol{\epsilon},\,{\bf 0}_3,\, \stackrel{k}{\overbrace{\boldsymbol{\epsilon},\dots,\boldsymbol{\epsilon}}},\,{\bf 0}_{n-2k})\,\,,\,\,\,\,\boldsymbol{\epsilon}=\left(\begin{matrix}0 & 1\cr -1 & 0\end{matrix}\right)\,.
\end{equation}
On the other hand the matrix ${\bf D}=-i\Gamma^1\Gamma^2/2$ is the spinorial representation of a generator $D$ which, in the fundamental representation of ${\rm SO}(5,n)$, has the following block-diagonal form:
\begin{equation}
D=i\,{\rm diag}(\boldsymbol{\epsilon},\,{\bf 0}_3,\, {\bf 0}_{2k},\,{\bf 0}_{n-2k})\,.
\end{equation}
This matrix can be written as follows
\begin{equation}
D=\frac{i}{k+1}\,J_0+J\,,\label{DJJ0}
\end{equation}
where $i\,J$ is the K\"ahler ${\rm U}(1)$-generator of the moduli space ${\rm SU(1,k)/S[U(1)\times U(k)]}$, so that $J$ is the pseudo-K\"ahler ${\rm O}(1,1)$-generator of the Wick-rotated space ${\rm SL(1+k)/GL(k)}$. The explicit form of $J$ in the fundamental representation of ${\rm SL(1+k)}$ is
given in Eq. (\ref{Jgen}). Equation (\ref{DJJ0}) implies that $D$ differs from $J$ by a matrix which is proportional to $J_0$ and which therefore commutes with ${\rm SL(1+k)}$. In particular $D$ and $J$ have the same eigenmatrices $N^\pm_\alpha$, see Appendix \ref{geodesiccharges}. Alternatively $i\,D$ can be viewed as the projection of $J_0$ on the subspace corresponding to the ${\bf 5}$, and labelled by the index $m$, of the R-symmetry group. This explains why the matrix $A_{1\,i}{}^j$, which should commute with the gauge group generators, is proportional in the background to the projection on the corresponding R-symmetry representation, of $J_0$, and thus, in the Euclidean theory, to ${\bf D}$.
With respect to ${\bf D}$ the spinorial representation of the Euclidean R-symmetry group decomposes as follows:
\begin{equation}
{\bf 4}\rightarrow {\bf 2}_{+\frac{1}{2}}+{\bf 2}_{-\frac{1}{2}}\,,
\end{equation}
where ${\bf 2}$ is the spinor representation of the ${\rm SU}(2)$ group commuting with ${\bf D}$.\par
The grading structure relative to ${\bf D}$, which we found for the shift matrices ${\bf N}^a$ in the extremal case, reflects the general structure of the Noether charge matrix $Q$ as expressed in \eqref{QNoether} in terms of the $N^\pm_\alpha$ nilpotent matrices. In particular in the $Q^2$ orbit $Q$ has a definite grading with respect to $J$, and thus to $D$, and this amounts to the fact that the ${\bf N}^a$ tensors have all the same gradings with respect to ${\bf D}$.\par
In the $Q^2=0$ orbit the Killing spinors are defined by the ${\bf 2}_{+\frac{1}{2}}$ representation if $\vec{m}=0$ and by the ${\bf 2}_{-\frac{1}{2}}$ representation if $\vec{p}=0$.

\section{Non-extremal solutions}\label{sec:non-extremal}
The extremal instantons described sofar  correspond to lightlike geodesics ($G_{IJ}\dot{\phi}^I\dot{\phi}^J = 0$). In this section we turn to  non-extremal instantons which therefore are defined by a non-zero geodesic velocity squared $G_{IJ}\dot{\phi}^I\dot{\phi}^J= c\neq 0$\,.
We explained already in section \ref{sec:general} that the sign of $c$ determines the qualitative features of the instantons.

If $c>0$ the solutions are under-extremal and correspond to a deformed EAdS metric that has a spike-like singularity in the middle. That singularity can potentially be interpreted as the position of the instanton and if so, we speculate the singularity gets resolved in full string theory\footnote{One feature of these solutions is that they can be rotated using the global symmetry $\SU(1,k)$ into a solution without axion fields. Such solutions can be consistently Wick-rotated to real solutions in Lorentzian AdS where they describe analogs of the singular flow found long time ago by Gubser \cite{Gubser:1999pk}. Unfortunately no clear holographic dual to that flow exists and it is yet unclear whether the singularity is physical since it does not pass some simple criteria \cite{Gubser:2000nd}.}

If $c<0$ the metric is a smooth double-sided Euclidean wormhole and the corresponding instanton could be called ``over-extremal''. Its existence is sometimes argued via the Weak Gravity Conjecture \cite{Hebecker:2016dsw, Montero:2015ofa}. Typically such wormhole solutions have singular scalar field profiles that are considered unphysical. Interestingly, a subset of the family of the geodesics of the sigma models considered here were  recently shown to be fully regular \cite{Hertog:2017owm} and this can be explicitly verified from the expressions we present below.

It is the aim of this section to provide the explicit expressions for the geodesics, discuss their orbit structure under the global symmetry group and to compute their on-shell action. For the latter we can use the boundary integral (\ref{onshellasboundary}) on the condition that the solution is regular in the bulk. This is problematic for the $c>0$ solutions. But we adopt the pragmatic attitude that the singularity will be resolved in full string theory such that we do not pick up a contribution in the on-shell action from the singular region. This approach at least gave sensible results in flat space with a single axion-dilaton pair \cite{Bergshoeff:2004pg}, where the on-shell action of the $c>0$ instanton correctly matched the mass of a non-extremal Reissner-Nordstr\"om black hole obtained from ``oxidising'' the instaton over the time-direction. For the wormholes $c<0$ the boundary formula now needs to evaluated at the left and right boundary of the wormhole and both contributions come with a relative minus sign.

\subsection{The general solutions}
One can readily check that the exponential of the charge matrix $Q$ (\ref{QNoether}) is given by
\begin{align}
\exp(2\tau Q)={\bf 1}+\frac{1}{\mu^2}\,Q^2 (\cosh(2\mu \tau)-1)+\frac{1}{\mu}\,Q \sinh(2\mu\tau )\quad\label{expQne},
\end{align}
where, $\mu=\sqrt{|\vec{m}\cdot \vec{p}|}$ and $c=4 \mu^2>0$. If $c=-4\mu^2<0$ we simply replace $\mu\rightarrow i\,\mu$.

Considering the right hand side of (\ref{equ4}) as in (\ref{expQne}), we obtain the following general solution for $\vec{m}\cdot \vec{p}>0$
\begin{align}
U\,&=\,\frac{1}{2}\,\log\left[\frac{\mu^2}{\left(m_0 \sinh\left(\mu \tau\right)-\mu\cosh\left(\mu \tau\right)\right)\left(p_0 \sinh\left(\mu \tau\right)-\mu\cosh\left(\mu \tau\right)\right)}\right]\quad,\label{gensolnex1}\\
\zeta^i\,&=\,\frac{1}{\sqrt{2}}\,\left[\,\frac{m_i}{m_0-\mu\coth\left(\mu \tau\right)}\,+\,\frac{p_i}{p_0-\mu\coth\left(\mu \tau\right)}\,\right]\quad,\label{gensolnex2}\\
\tilde{\zeta}_i\,&=\,\frac{1}{\sqrt{2}}\,\left[\,-\frac{m_i}{m_0-\mu\coth\left(\mu \tau\right)}\,+\,\frac{p_i}{p_0-\mu\coth\left(\mu \tau\right)}\,\right]\quad,\label{gensolnex3}\\
a\,&=\,-\frac{m_0}{m_0-\mu\coth\left(\mu \tau\right)}\,+\,\frac{p_0}{p_0-\mu\coth\left(\mu \tau\right)}\quad\label{gensolnex4},
\end{align}
 where $i=1,\dots, k-1$, as before.\\
If $\vec{m}\cdot \vec{p}<0$ the solution can be rewritten in the following form:
\begin{align}
U\,&=\,\frac{1}{2}\,\log\left[\frac{\mu^2}{\left(m_0 \sin\left(\mu \tau\right)-\mu\cos\left(\mu \tau\right)\right)\left(p_0 \sin\left(\mu \tau\right)-\mu\cos\left(\mu \tau\right)\right)}\right]\quad,\label{gensolnex1}\\
\zeta^i\,&=\,\frac{1}{\sqrt{2}}\,\left[\,\frac{m_i}{m_0-\mu\,{\rm cotg}\left(\mu \tau\right)}\,+\,\frac{p_i}{p_0-\mu\,{\rm cotg}\left(\mu \tau\right)}\,\right]\quad,\label{gensolnex2}\\
\tilde{\zeta}_i\,&=\,\frac{1}{\sqrt{2}}\,\left[\,-\frac{m_i}{m_0-\mu\,{\rm cotg}\left(\mu \tau\right)}\,+\,\frac{p_i}{p_0-\mu\,{\rm cotg}\left(\mu \tau\right)}\,\right]\quad,\label{gensolnex3}\\
a\,&=\,-\frac{m_0}{m_0-\mu\,{\rm cotg}\left(\mu \tau\right)}\,+\,\frac{p_0}{p_0-\mu\,{\rm cotg}\left(\mu \tau \right)}\quad\label{gensolnex5}.
\end{align}
Similar to the lightlike geodesic curves, we can use global shift-like symmetries (\ref{shifstsym}) to find the general curve that does not pass through the origin at $\tau=0$. We can also rotate the solutions through the origin into simple ``generating solutions'' desribed by the normal form of $Q$ under $\SO(k)$. This gives us the orbit structure. As before this implies that $\vec{p}=(p_0,0,\dots,0)$ and $\vec{m}=(m_0,m_1,0,\dots,0)$. There is now only one orbit for each sign of $c$:
\begin{align}
& c>0:\quad  p_0m_0>0\,,\nonumber\\
& c<0:\quad p_0m_0<0\,.
\end{align}

\subsection{On-shell action}
We now compute the on-shell action for the non-extremal solutions with $c>0$. We evaluate the boundary action using the formula (\ref{onshellasboundary}), and find:
\begin{equation}
S_{{\rm on-shell}}^{\rm real}=\frac{Vol(S^4)}{2\,\kappa^2_5}\,\frac{1}{\hat{m}^2_0\hat{p}^2_0 }\,{\rm Abs}\left[\frac{(\hat{m}_0+\hat{p}_0)}{2}\,\sum_{i=1}^{k-1}\,\left(\hat{m}_0\,p_i-\hat{p}_0\,m_i\right)^2- \mu\,\hat{m}_0\hat{p}_0({m}_0-{p}_0) \right]\,,\label{Sbdryne}
\end{equation}
where we have defined:
\begin{equation}
\hat{m}_0=m_0-\mu\,,\,\,\,\,\hat{p}_0=p_0-\mu\,\,,\,\,\,\,\,\mu=\sqrt{\vec{m}\cdot\vec{p}}\,.
\end{equation}
The reader can verify that in the extremal limit $\mu\rightarrow 0$, the expression (\ref{Sbdryne}) reduces to the corresponding formula for the extremal case.

In case $c<0$, the on-shell action is similar but now involves the subtraction of the boundary term $\Pi$ on the left and on the right side of the wormhole and we leave a detailed discussion of these on-shell actions for a separate work \cite{inprogress}.

\section{Discussion}\label{sec:discussion}
Let us summarize the results of this paper. We have argued that instanton solutions of IIB supergravity in Euclidean $\rm AdS_5\times S^5/\mathbb{Z}_k$ are completely characterized by the geodesic curves in the moduli space of the Euclidean vacuum, $\mathcal{M}_{\text{moduli}} = \SL(k+1,\mathbb{R})/\GL(k,\mathbb{R})$, which is a suitable Wick-rotation of the moduli-space of the Lorentzian vacuum: $\mathcal{M}_{\text{moduli}} = \SU(1,k)/\rm S[U(1)\times U(k)]$. We have found the explicit expression for the general geodesic curve and computed the on-shell action in terms of the charges.

Our main focus was on the extremal instanton solutions given by the lightlike geodesics. The metric then remains pure Euclidean AdS since the energy-momentum tensor vanishes. The lightlike geodesics are separated into two classes depending on the nilpotency of the Noether charge matrix $Q$. If $Q^2=0$ the solutions preserve 8 out of the original 16 supercharges. The remaining lightlike geodesics have $Q^3=0$ and break all supersymmetries.

An obvious question for future research is the map between these instantons and the instantons of the holographic dual 4D $\mathcal{N}=2$ necklace quiver theories \cite{Kachru:1998ys, Corrado:2002wx}. Especially for the supersymmetric solutions it is tempting to expect that a detailed correspondence should work out and we hope to come back  to this in a future work. Some relevant studies of instantons of the necklace quivers can be found in \cite{Hollowood:1999bm, Argurio:2012iw}. If the extremality condition in the supergravity condition maps to the self-duality of the Yang--Mills field strengths then our results suggest that the dual quiver gauge theories should have a whole zoo of non-supersymmetric but self-dual solutions dual to the sugra solutions with $Q^3=0$.

Closely related to the gauge theory dual interpretation is the stringy interpretation of these instantons. Since the massless fields consist of the axio-dilaton in IIB and $2(k-1)$ fields corresponding to the periods of $B_2$ and $C_2$ over the shrinking two-cycles of $\rm S^5/\mathbb{Z}_k$ (twisted sector) the uplift to 10D should correspond to a mixture of the standard D-instanton and various fractional  D-instantons. The fractional  D-instantons can be regarded as Euclidean F1 and D1 strings wrapping the shrinking cycles. Hence we expect the $m_0$ ($p_0$) charges to originate from D-instantons sources and the $m_i$ ($p_i$) charges to originate from the Euclidean D1's (F1's) wrapping the vanishing two-cycles. The fact that the latter charges can be fractional seems consistent with the fractional contributions $m_i^2/m_0$ in the on-shell action (\ref{boundintegralBPS}).

Finally we note that the observation that supergravity instantons are geodesics on the AdS moduli-space is of course not restricted to $\rm AdS_5\times S^5/\mathbb{Z}_k$ and this should hold in general. The holographic correspondence between an AdS moduli space and the conformal manifold of the dual field theory then suggests the general result that \emph{geodesics on the conformal manifold are in  correspondence with instantons of the CFT at large N}. If correct, this is rather intriguing, since it is far from obvious how the solutions of the self-duality equation should know about the Zamolodchikov metric on the conformal manifold.

Therefore a natural extension of this work would be the investigation of instantons in Euclidean $\rm AdS_3 \times S^3 \times T^4$ or $\rm AdS_3 \times S^3 \times K3$. The dual (D1-D5) CFT's have conformal manifolds of the type \cite{Cecotti:1990kz}
\be
\frac{\SO(4,n)}{\SO(4)\times \SO(n)}\,,
\ee
with $n=20$ for $K3$. However the analogues supergravity analysis of the AdS moduli spaces in $D>3$ \cite{Corrado:2002wx, deAlwis:2013jaa, Louis:2015dca, Louis:2016msm, Louis:2016qca, Louis:2015mka, Lust:2017aqj} has not been carried out in 3D gauged supergravity.

\section*{Acknowledgements}
We thank Riccardo Argurio, Matteo Bertolini, Nikolay Bobev, Hagen Triendl and Jakob Moritz for useful discussions. The work of  TVR is supported by the FWO odysseus grant G.0.E52.14N and by the C16/16/005 grant of the KULeuven. We furthermore acknowledge support from the European Science Foundation Holograv Network  and the COST Action MP1210 `The String Theory Universe'.

\newpage
\appendix
\section{A note on parametrizations of ${\rm EAdS}_5$}\label{A1}
The two parametrizations of ${\rm EAdS}_5$ that we refer to in this work are $x^\mu=(z,\vec{x})$, $\vec{x}=(x^1,\dots, x^4)$, and $x^\mu=(r,\phi^\ell)$, $\ell =1,\dots,4$ in which the metric reads:
\begin{equation}
ds^2=\frac{\ell^2}{z^2}\left(dz^2+|d\vec{x}|^2\right)=\frac{dr^2}{1+\frac{r^2}{\ell^2}}+r^2\,d^2\Omega(S_4)\,,
\end{equation}
where in the \emph{radial parametrization} $(r,\phi^\ell)$, $\phi^\ell$ parametrize a 4-sphere $S_4$ of unit radius, whose line element is denoted by $d^2\Omega(S_4)$. The radial variable $r$, as a function of $z,\,\vec{x}$ is given by:
\begin{equation}
r(z,\,\vec{x})=F(z,\,\vec{x})\,,
\end{equation}
where $F(z,\,\vec{x})$ is given in (\ref{FFunc}) with $\vec{x}_0=\vec{0},\,z_0=\ell$.\par
In the radial parametrization, if $H(r)$ is the spherically symmetric harmonic function in Eq. (\ref{Hharm}), we have the following useful formula:
\begin{equation}
\sqrt{|g_5|}\,g^{rr}\,\partial_r H=-3\,\alpha\, \sqrt{|g(S_4)|}\,,
\end{equation}
where $|g(S_4)|$ is the determinant of the metric on the unit $S_4$.
This relation is useful when computing the integral over  ${\rm EAdS}_5$ of a Lagrangian density evaluated on solutions which only depend on $H$.
We conveniently choose $\alpha=1/3$.\par
The boundary of ${\rm EAdS}_5$ is located at $r\rightarrow\infty$ which corresponds to $z=0$. The parameter $\beta$ in (\ref{Hharm}) is fixed requiring that $H=0$ at the boundary.

\section{The coset construction}\label{A2}
We consider the scalar manifold
\begin{equation}\label{coset1}
\mathcal{M}_{\text{moduli}}=\frac{\text{SL}\left(k+1\right)}{\text{GL}\left(k\right)}\quad,
\end{equation}
which is conveniently described in terms of a solvable Lie algebra parametrization, in which the scalar manifold $ \mathcal{M}_{\text{moduli}}$ is globally described as isometric to a solvable group manifold generated by $Solv\text{  :  }\mathcal{M}_{\text{moduli}}\text{ }\thicksim\exp(Solv)$. The scalar fields $U, \zeta^i, \tilde{\zeta}_i, a$ parametrize respectively the generators $H_0, T_i, T^i, T_{\bullet}$ of $Solv$ via the coset representative
\begin{equation}
\mathbb{L} = \exp(-a T_{\bullet})\exp(\sqrt{2}\mathcal{Z}^M T_{M})\exp(2U H_{0})\,,
\end{equation}
where  $\mathcal{Z}^M\equiv\left(\zeta^i\text{,}\tilde{\zeta}_i\right)$. The index $i$ runs over $1\ldots k-1$.
The solvable generators have the explicit form
\begin{equation}
\begin{split}
H_0&=\frac{1}{2}\left(e_{1,k+1}+e_{k+1,k}\right)\quad,\\
T_{i}^{(1)}=T_{i}&=\frac{1}{2}\left(e_{i+1,k+1}-e_{k+1,i+1}-e_{1,i+1}-e_{i+1,1}\right)\quad,\\
T_{i}^{(2)}=T^{i}&=\frac{1}{2}\left(e_{1,i+1}+e_{k+1,i+1}+e_{i+1,k+1}-e_{i+1,1}\right)\quad,\\
T_{\bullet}&=\frac{1}{2}\left(e_{1,1}+e_{k+1,1}-e_{1,k+1}-e_{k+1,k+1}\right)\,.\label{solvgen}
\end{split}
\end{equation}
From the solvable generators in (\ref{solvgen}) one can construct the following $2k$ non-compact generators
\begin{equation}\label{embedding}
\begin{split}
K_0&=H_0\quad,\\
K_{i}^{(1)}&=\frac{1}{2}\left(T_{i}^{(1)}+{T_{i}^{(1)}}^{T}\right)\quad,\\
K_{i}^{(2)}&=\frac{1}{2}\left(T_{i}^{(2)}-{T_{i}^{(2)}}^{T}\right)\quad,\\
K_{\bullet}&=\frac{1}{2}\left(T_{\bullet}-{T_{\bullet}}^{T}\right)\,.
\end{split}\,
\end{equation}
The isometry algebra $\mathfrak{g}=\mathfrak{sl}(k+1)$ splits into the isotropy algebra $\mathfrak{H}=\mathfrak{gl}(k)$ and the coset space $\mathfrak{K}$. The Cartan involution $\sigma$ leaving $\mathfrak{H}$ invariant acts as $\sigma(g)=\eta\,g\,\eta$, where $\eta$ is the
$\text{GL}\left(k\right)$-invariant matrix
\begin{equation}
\eta=\left(
                                                                                \begin{array}{cccc}
                                                                                  1 & 0 & \cdots & 0 \\
                                                                                  0 & -1 &  & \vdots \\
                                                                                  \vdots &  & \ddots & 0 \\
                                                                                  0 & \cdots & 0 & -1 \\
                                                                                \end{array}
                                                                              \right)\quad.
\end{equation}
The matrix $M(\phi)$, defined as,
\begin{equation}
M(\phi)\,=\,\mathbb{L}\eta\mathbb{L}^{-1}\eta\,,\label{equ3}
\end{equation}
is manifestly invariant under $\mathbb{L}\rightarrow \mathbb{L}\,h$, where $h\in {\rm GL}(k)$. We also have that
\begin{equation}
M^{-1}\d M=2 \sigma(\mathbb{L})\,P\,\sigma(\mathbb{L}^{-1})\,,
\end{equation}
where $P$ is the vielbein 1-form matrix. Then the metric can be written in the form
\begin{equation}
G_{IJ}(\phi)=\frac{1}{2}\, {\rm Tr}(M^{-1}\partial_I M M^{-1}\partial_J M)\,,
\end{equation}
and leads to the expression (\ref{metric}).

The explicit embedding of the moduli-space coset (\ref{coset2}) (or \ref{coset1}) into the bigger coset (\ref{coset3}) of half-maximal supergravity is necessary for computing the matrices $\mathcal{V}$ used in the analysis of the Killing-spinor equations. The explicit embedding of the  ${\rm SL}(k+1)$ Lie algebra generators, inside ${\rm SO}(5,n)$ solvable generators is given by
\begin{equation}\label{solvgenSO5n}
\begin{split}
H_0&=\tfrac{1}{2}\left(e_{1,2k+4}+e_{2,2k+5}+e_{2k+4,1}+e_{2k+5,2}\right)\,,\\
T_{i}&=-\tfrac{1}{2}\left(e_{1,2i+4}+e_{2,2i+5}+e_{2i+4,1}+e_{2i+5,2}-e_{2i+4,2k+4}-e_{2i+5,2k+5}+e_{2k+4,2i+4}+e_{2k+5,2i+5}\right)\,,\\
-i\,T^{i}&=-\tfrac{1}{2}\left(e_{1,2i+5}-e_{2,2i+4}+e_{2i+5,1}-e_{2i+4,2}+e_{2i+4,2k+5}-e_{2i+5,2k+4}+e_{2k+4,2i+5}-e_{2k+5,2i+4}\right)\,,\\
-i\,T_{\bullet}&=-\tfrac{1}{2}\left(e_{1,2}-e_{2,1}-e_{1,2k+5}+e_{2,2k+4}+e_{2k+4,2}-
e_{2k+5,1}-e_{2k+4,2k+5}+e_{2k+5,2k+4}\right)\,,
\end{split}
\end{equation}
where the $-i$ factors in the left hand sides of the last two equations are due to the fact that $T^{i},\,T_{\bullet}$ are ${\rm SL}(k+1)$-generators, so that $-i\,T^{i},\,-i\,T_{\bullet}$ are the
${\rm SU}(1,k)$-generators embedded in ${\rm SO}(5,n)$ as described in \cite{Louis:2015dca}.

\section{The geodesic charges}\label{geodesiccharges}
The following generators
\begin{equation}
\begin{split}
N^{\pm}_i&=-\left(K_{i}^{(1)}{\pm}K_{i}^{(2)}\right)\,,\\
N^{\pm}_{0}&=N^{\pm}_{\bullet}=K_0{\pm}K_{\bullet}\quad\label{nilpgen}\,,\\
\end{split}
\end{equation}
are all nilpotent and the corresponding matrices $N^{\pm}_{\alpha}$, $\alpha=0,\dots, k-1$, have definite gradings with respect to the ${\rm SO}(1,1)$ generator $J$ of the pseudo-K\"ahler transformations
\begin{equation}
J=\frac{1}{k+1}{\rm diag}(-k,+1,+1,\dots,+1)\,,\label{Jgen}
\end{equation}
which commutes with the $\mathfrak{sl}(k)$ subalgebra of $\mathfrak{H}$.
One can indeed verify that
\begin{equation}
[J,\,N^{\pm}_{\alpha}]=\pm N^{\pm}_{\alpha}\,.
\end{equation}
The solution $\hat{\phi}^I(\tau)$ to the geodesic equation, defined by the values of the scalar fields at radial infinity $\hat{\phi}^I(\tau=0)=\phi^I_0$ and Noether matrix $\hat{Q}$, can be written as the solution to the matrix equation
\begin{equation}
M(\hat{\phi}^(\tau))\,=M(\phi_0)\,\exp(2\,\hat{Q}\, \tau)\,, \label{equ4}
\end{equation}
where $\phi_0\equiv (\phi_0^I)$. It can be obtained from a geodesic ${\phi}^I(\tau)$ with initial point ${\phi}^I(\tau=0)=0$ and Noether matrix ${Q}$ through the transformation $\mathbb{L}_0\equiv \mathbb{L}(\phi_0)$:
\begin{equation}
M(\hat{\phi}(\tau))=\mathbb{L}_0\,M({\phi}(\tau))\sigma(\mathbb{L}_0)^{-1}\,\,,\,\,\,\,\,\hat{Q}=\sigma(\mathbb{L}_0)
{Q}\sigma(\mathbb{L}_0)^{-1}\,.
\end{equation}
Let us concentrate on the solution through the origin. The corresponding Noether matrix belongs to the coset space $Q\in \mathfrak{sl}(k+1)\ominus \mathfrak{gl}(k)$ and can be expressed as the following linear combination of the coset generators in (\ref{nilpgen})
\begin{equation}
Q\,=\,\sum_{\alpha=0}^{k-1}\,\left(\,p_{\alpha}\,N^{+}_{\alpha}\,+\,m_{\alpha}\,N^{-}_{\alpha}\,\right)\,=\,\left(
                                                                                                \begin{array}{c|cccc}
                                                                                                  0 & m_1 & \cdots & m_{k-1}& m_0 \\ \hline
                                                                                                 p_1 & 0 & \cdots & 0& 0 \\
                                                                                                  \vdots & \vdots & \ddots & & \vdots \\
                                                                                                  p_{k-1} & 0 & \cdots & 0 & 0 \\
                                                                                                  p_{0} & 0 & \cdots & 0 & 0 \\
                                                                                                \end{array}
                                                                                              \right)\,.\label{QNoether}
\end{equation}
The total geodesic velocity squared, is then given by the simple inner-product:
\begin{equation}
c=2 {\rm Tr}(Q^2)=4\,\vec{p}\cdot \vec{m}\,,
\end{equation}
where $\vec{m}=\left(m_0,\dots,m_{k-1}\right)$ and $\vec{p}=\left(p_0,\dots,p_{k-1}\right)$.
The nilpotency condition for $Q$ is:
\begin{equation}
\mbox{ Q nilpotent}\,\,\Leftrightarrow\,\,\,\, \vec{m}\cdot \vec{p}=0\,.\label{nilpot}
\end{equation}
In this case there are two nilpotent orbits:
\begin{itemize}
\item{The orbit of degree $2$ ($Q^2=0$), obtained when all coefficient $p$ or $m$ are zero. In this case $Q$ has a definite grading with respect to the pseudo-K\"ahler generator $J$:
    \begin{equation}
    [J,\,Q]=Q\,\,(\vec{m}=0)\,\,\,;\,\,\,\,\,\,    [J,\,Q]=-Q\,\,(\vec{p}=0)\,;\label{gradingQ}
    \end{equation}
     }
\item {The orbit of degree $3$ ($Q^3=0$) otherwise.}
\end{itemize}
The grading property of $Q$ in the first class has a bearing as to the supersymmetry properties of the corresponding solutions, as explained in Section \ref{sec:susy}.

The equation (\ref{equ4}) admits the general solution presented in the main text in equations (\ref{gensol1}-\ref{gensol4}).

\section{Some simple solutions} \label{k=1}

When $k=1$  $\zeta^i=\tilde{\zeta}_i=0$. If we call $\phi=- 2 U$ and $\chi=a$, to make contact with \cite{Bergshoeff:2005zf}, we find
from (\ref{gensol1}-\ref{gensol4}):
\begin{equation}
e^\phi= (1-p \tau) (1-m \tau)\,,\,\,\,\chi=\frac{1}{1-\tau m}-\frac{1}{1-\tau p}\,.
\end{equation}
We can either set $m=0$ or $p=0$. Regularity requires $(1-p \tau)(1-m \tau)>0$.
\paragraph{The anti-instanton.}
Setting $m=0$ and $$1-p \tau=|q_-|\,{H}\,,$$ where the harmonic function ${H}$ is the one used in  \cite{Bergshoeff:2005zf},
we have
$$e^\phi=|q_-|\,{H}\,\,,\,\,\,\,\chi=1-\frac{1}{|q_-|\,{H}}\,.$$
If we define  $q_-=-|q_-|$ and shift $\chi\rightarrow \chi-1+\frac{q_3}{q_-}$, we end up with the \emph{anti-instanton} solution of
\cite{Bergshoeff:2005zf}.\par
\paragraph{The instanton.} If we set $p=0$, $$1-m \tau=|q_-|\,{H}\,,$$ and $q_-=|q_-|>0$ we get:
$$e^\phi=|q_-|\,\tau\,\,,\,\,\,\,\chi=\frac{1}{q_-\,{H}}-1\,.$$
Shifting $\chi\rightarrow \chi+1+\frac{q_3}{q_-}$ we end up with the \emph{instanton} solution of
\cite{Bergshoeff:2005zf}.

\section{Clifford algebra of ${\rm SO}(5)$}\label{Cliff}
The gamma-matrices $(\Gamma^m)_i{}^j$, $i,j=1,\dots,4$, of ${\rm SO}(5)$ are $4\times 4$ matrices satisfying:
\begin{equation}
\{\Gamma^m,\,\Gamma^n\}=\Gamma^m\,\Gamma^n+\Gamma^n\,\Gamma^m=2\delta^{mn}\,{\bf 1}_4\,\,,\,\,\,m,n=1,\dots,5\,,
\end{equation}
where ${\bf 1}_4$ is the $4\times 4$ identity matrix. The spinorial representation of ${\rm SO}(5)$
is the fundamental representation of ${\rm USp}(4)$ and features an antisymmetric invariant matrix $\Omega^{ij}$, which coincides with the charge conjugation matrix $C^{ij}$, satisfying $C\Gamma^mC^{-1}=\Gamma^{m\,T}$:
\begin{equation}
\Omega^{ij}=C^{ij}\,\,,\,\,\,\,\Omega_{ij}\equiv \Omega^{ij}\,.
\end{equation}
The indices are lowered and raised using $\Omega_{ij}$ and $\Omega^{ij}$, respectively, using the North-West, South-East convention.
In particular we define the matrices:
\begin{equation}
(\Gamma_m)^{ij}=\Omega^{ik}\,(\Gamma_m)_k{}^j\,,
\end{equation}
which satisfy the properties:
\begin{equation}
(\Gamma_m)^{ij}=-(\Gamma_m)^{ji}\,\,,\,\,\,(\Gamma_m)^{ij}\Omega_{ij}=0\,\,,\,\,\,
(\Gamma_m)_{ij}=\Omega_{ik}\Omega_{jl}(\Gamma_m)^{kl}=((\Gamma_m)^{ij})^*\,.
\end{equation}
The antisymmetric couple $[ij]$ in $(\Gamma_m)^{ij}$ labels the representation ${\bf 5}$ of ${\rm USp}(4)$, described as the antisymmetric, traceless product of two ${\bf 4}$ representations, which also coincides with the fundamental representation of ${\rm SO}(5)$. The tensor $(\Gamma_m)^{ij}$ intertwines between the two different descriptions of the same representation.\par
We choose for them the following explicit representation:
\begin{align}
\Gamma^1&=\left(
\begin{array}{cccc}
 0 & 0 & 1 & 0 \\
 0 & 0 & 0 & 1 \\
 1 & 0 & 0 & 0 \\
 0 & 1 & 0 & 0
\end{array}
\right)=\sigma_1\times {\bf 1}_2\,\,,\,\,\,\,\,\Gamma^2=\left(
\begin{array}{cccc}
 0 & 0 & 0 & i \\
 0 & 0 & i & 0 \\
 0 & -i & 0 & 0 \\
 -i & 0 & 0 & 0
\end{array}
\right)=-\sigma_2\times \sigma_1\,,\nonumber\\
\Gamma^3&=\left(
\begin{array}{cccc}
 0 & 0 & 0 & 1 \\
 0 & 0 & -1 & 0 \\
 0 & -1 & 0 & 0 \\
 1 & 0 & 0 & 0
\end{array}
\right)=-\sigma_2\times \sigma_2\,\,,\,\,\,\,\,\Gamma^4=\left(
\begin{array}{cccc}
 0 & 0 & i & 0 \\
 0 & 0 & 0 & -i \\
 -i & 0 & 0 & 0 \\
 0 & i & 0 & 0
\end{array}
\right)=-\sigma_2\times \sigma_3\,,\nonumber\\
\Gamma^5&=\left(
\begin{array}{cccc}
 1 & 0 & 0 & 0 \\
 0 & 1 & 0 & 0 \\
 0 & 0 & -1 & 0 \\
 0 & 0 & 0 & -1
\end{array}
\right)=\sigma_3\times {\bf 1}_2=+\Gamma^1\Gamma^2\Gamma^3\Gamma^4\,,
\end{align}
where $\sigma_x$, $x=1,2,3$, are the usual Pauli matrices.
In this basis $\Omega=(\Omega^{ij})=C$ reads: $\Omega=\Gamma^4\Gamma^2={\bf 1}_2\times i\sigma_2$.
We refer to Appendix A of \cite{Louis:2015dca} for the properties of these matrices.\

\bibliographystyle{utphys}
{\small
\bibliography{refs}}

\end{document}